\documentclass[aps,preprint,superscriptaddress,bibliography,prl]{revtex4-1}

\usepackage{graphicx, amsmath, verbatim, dsfont, amsfonts, color, braket}
\usepackage{physics}
\usepackage[us]{datetime}

\begin{document}
\title{Disorder-generated non-Abelions}
\author{G.~Simion}
\affiliation{Department of Physics and Astronomy, Purdue University, West Lafayette, IN 47907 USA}
\author{A.~Kazakov}
\affiliation{Department of Physics and Astronomy, Purdue University, West Lafayette, IN 47907 USA}
\author{L.~P.~Rokhinson}
\affiliation{Department of Physics and Astronomy, Purdue University, West Lafayette, IN 47907 USA}
\affiliation{Birck Nanotechnology Center, Purdue University, West Lafayette, IN 47907 USA}
\affiliation{Department of Electrical and Computer Engineering, Purdue University, West Lafayette, IN 47907 USA}
\author{T.~Wojtowicz}
\affiliation{International Research Centre MagTop, Aleja Lotnikow 32/46, PL-02668 Warsaw, Poland}
\affiliation{Institute of Physics, Polish Academy of Sciences, Al. Lotnikow 32/46, 02-668 Warsaw, Poland}
\author{Y.~B.~Lyanda-Geller}
\email{yuli@purdue.edu}
\affiliation{Department of Physics and Astronomy, Purdue University, West Lafayette, IN 47907 USA}
\affiliation{Birck Nanotechnology Center, Purdue University, West Lafayette, IN 47907 USA}

\date{July, 12 2017}

\begin{abstract}
Two classes of topological superconductors and Majorana modes in condensed matter systems are known to date:
one, in which disorder induced by impurities strongly suppresses topological superconducting gap and is detrimental to Majorana modes, and the other, where Majorana fermions are protected by disorder-robust topological superconductor gap.
In this work we predict a third class of topological superconductivity and Majorana modes, in which  topological superconductivity and Majorana fermions appear exclusively in the presence of impurity disorder. Observation and control of Majorana fermions and other non-Abelions often requires a symmetry of an underlying system leading to a gap in a single-particle or quasiparticle spectra. Disorder introduces states into the gap and enables conductance and proximity-induced superconductivity via the in-gap states. We show that disorder-enabled topological superconductivity can be realized in a quantum Hall ferromagnet, when helical domain walls are coupled to an s-wave superconductor. 
Solving a general quantum mechanical problem of disorder-induced bound states in a system of spin-orbit coupled Landau levels, we demonstrate that disorder-induced Majorana modes emerge at phase boundaries in a specific case of a quantum Hall ferromagnetic transition in CdMnTe quantum wells  at a filling factor $\nu=2$. 
Recent experiments on transport through electrostatically controlled individual domain wall in this system indicated the vital role of disorder in conductance, but left an unresolved question whether this could intrinsically preclude generation of Majorana fermions and other non-Abelions. The proposed resolution of the problem demonstrating emergence of Majorana fermions exclusively due to impurity disorder in experimental setting of domain walls in quantum Hall ferromagnets opens a path forward. We show that electrostatic control of domain walls in an integer quantum Hall ferromagnet allows manipulation of Majorana modes. Similar physics can possibly emerge for ferromagnetic transitions in the fractional quantum Hall regime leading to the formation and control of higher order non-Abelian excitations.
\end{abstract}

\maketitle

\section{Introduction}

Non-abelions in solid state systems, such as Majorana fermions, parafermions or Fibonacci anyons,
result in topologically degenerate ground state characterized by non-Abelian statistics and provide paths to topological fault-tolerant quantum computing \cite{Kitaev2003,Nayak2008}. Exotic states with non-Abelian excitations are predicted to emerge in correlated states in the fractional quantum Hall regime in two-dimensional electron, bilayer and hole gases \cite{MooreRead, Review, ReadRezayiPRB99, BarkeshliWenNATPRL, PetersonDasSarmaPRB, PapicGoerbig12PRB, Simion2017}, in $p$-wave $^3$He \cite{Volovik}, and in hybrid superconductor/topological insulator \cite{FuKanePRLMajorana,Molenkamp2017} and superconductor/semiconductor \cite{DasSarmaMajoranaPRL10,Oreg10,Mourik2012,Rokhinson2012a} systems. Topological superconductors can be divided into two broad classes: one, in which disorder induced by impurities strongly suppresses topological superconducting gap and can be detrimental to non-Abelions \cite{Lee2011,Huse2001,Stanescu2011,Lutchin2011,Brouwer2011,Brouwer2011L,Sau2012}, and the other, in which non-Abelian excitations are protected by a disorder-robust topological superconductor gap \cite{dasarma_chain_2012,Akhmerov_chain_2013,Vishveshwara,Franz,Nori}. In this work, we present a third class of topological superconductivity  and Majorana fermions, which appear exclusively in the presence of disorder within an otherwise gapped energy spectrum.

Observation and control of Majorana fermions and other non-Abelions often requires a symmetry of an underlying system leading to a gap in a single-particle or a quasiparticle spectra. An example is a quantum Hall system proximity-coupled to a superconductor, where Majorana fermions \cite{SternBerg2012}, parafermions \cite{Clarke2012}, and Fibonacci fermions \cite{Mong2013} are predicted to be formed in the presence of interacting counter-propagating edge channels. Experiments, though, indicate strong level repulsion and opening of a large exchange gap for interacting edge channels with the same orbital quantum numbers\cite{Kazakov2016}.  A promising
alternative is a quantum Hall ferromagnetic transition, where coupled counter-propagating chiral states at the boundaries of ferromagnetic domains form helical domain walls \cite{falko00}, perfect precursors for the formation of topological channels in the presence of superconducting interactions.
However, even when helical domain walls are formed from almost orthogonal states with different orbital quantum numbers and opposite spins in integer and fractional QHE regimes, spin-orbit interactions open small spectral gaps in bulk Landau level spectrum \cite{Kazakov2016} and in the spectra of electrostatically induced edge states \cite{Kazakov2017}. These gaps suppress electron transport at low temperatures \cite{eisenstein90,Kazakov2016}, but  in short helical domain walls  transport can be carried by the in-gap states \cite{Kazakov2017}.

Here we demonstrate that disorder-induced in-gap states in electrostatically defined helical domain walls can lead to topological  superconductivity when coupled to an s-wave superconductor. We solve a general quantum mechanical problem of impurity states in the presence of Landau quantization and spin-orbit interactions, and derive impurity states in the electrostatically induced domain wall.  We then map the system of these states on the generalized Kitaev chain \cite{kitaev_unpaired_2001,dasarma_chain_2012}, and calculate the phase diagram for the existence of  topological superconductivity and Majorana fermions. Finally, we demonstrate that with a local control of the QHFm transition it is possible to induce, move, exchange, fuse and braid Majorana modes.

We consider a specific case of a 2D electron gas formed in asymmetrically Mn-doped CdTe quantum wells, where local electrostatic control of a quantum Hall ferromagnetic transition and a single helical domain wall manipulation have been recently reported\cite{Kazakov2017}. However, these experiments  indicated the vital role of disorder in conductance through electrostatically controlled individual domain wall in this system, and raised the question whether this intrinsically precludes generation of Majorana fermions and other non-Abelions in quantum Hall ferromagnets or non-Abelions are still feasible. We resolve this problem affirmatively, showing that disorder is crucial for generating non-Abelions.
Our conclusions should be applicable to any system where quantum Hall ferromagnetic transition can be locally controlled, such as, e.g., a 2D hole gas in Ge\cite{Lu2017}. Quantum Hall ferromagnetic transitions in the integer and fractional QHE regimes have been observed in 2D gases in many semiconductors, including GaAs \cite{eisenstein90,Koch1993}, AlAs \cite{DePoortere2000}, InSb \cite{Chokomakoua2004}, CdMnTe\cite{Jaroszynski2002,Betthausen2014}, Si \cite{Lai2006} and graphene\cite{Feldman2013}, and their electrostatic control has been shown \cite{Klitzing2002,Kazakov2016}.

\section{ Majorana modes in a helical domain wall}

\begin{figure}[h]
\includegraphics[width=0.9\textwidth]{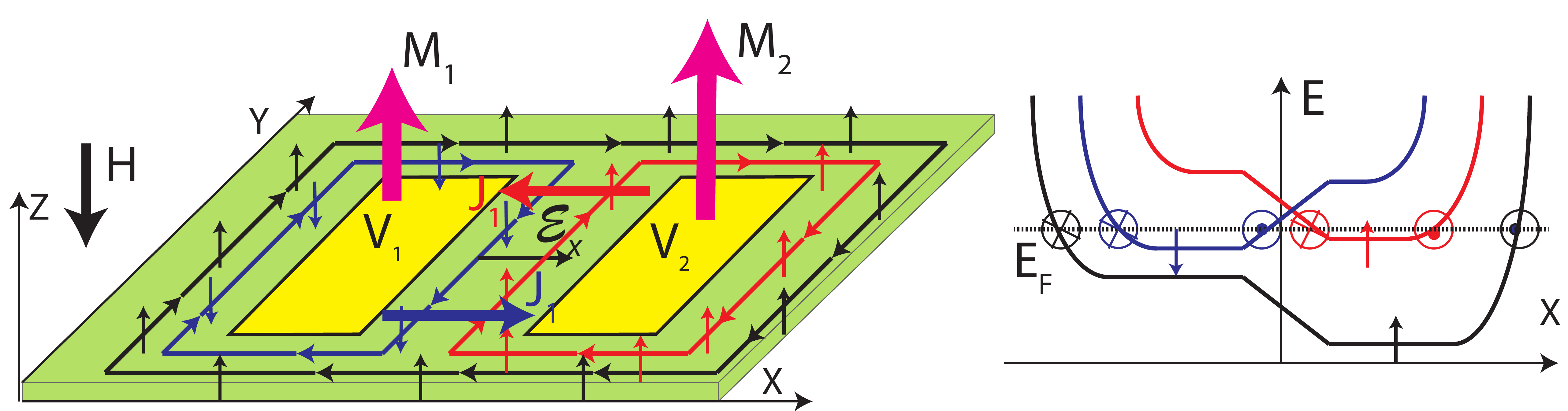}
\caption{a) Electrostatic gates $V_1$ and $V_2$ control magnetization $\bf{M}_1$ and $\bf{M}_2$  caused by the electron exchange interactions with Mn impurities. A spatial gradient of magnetization $J_1$ and a potential gradient $\mathcal E_x$ result in the formation of edge-like states between the gates (red and blue). Vertical arrows along edge states show spin polarization of electrons, which is opposite for edge-like states as a result of quantum Hall ferromagnetic transition. Between the gates edge-like states have opposite  velocities. Hybridized, they form a helical domain wall. b) Energy profile of electron states in the absence of spin-orbit interactions. Due to different polarization of red and blue states, the electron system at $\nu=2$, which also include electrons in the ground Landau level  (black), are unpolarized on the left and polarized on the right.}
\label{fig:varg_edges}
\end{figure}

In Mn-doped CdTe quantum wells external magnetic field $B$ aligns spins of Mn$^{2+}$ ions and generates an additional exchange contribution to the electron spin splitting due to interactions between conduction electrons and d-shell electrons localized on Mn\cite{Wojtowicz1999}. This s-d exchange splitting has a sign opposite to the bare Zeeman splitting for electrons in the conduction band, leading to multiple level crossings at high magnetic fields\cite{Jaroszynski2002}. The ferromagnetic transition of interest occurs at a crossing of states with opposite polarizations belonging to the first two Landau levels $(n=0,\uparrow)$ and $(n=1,\downarrow)$ at a filling factor $\nu=2$. In asymmetrically Mn-doped quantum wells the strength of the s-d exchange can be electrostatically controlled \cite{Kazakov2016} and it is possible to form an unpolarized and a fully polarized states under different gates,as shown schematically in Fig.~\ref{fig:varg_edges}.

In order to describe a helical domain wall formed between the gates we consider the edge-like states in a quantum Hall system induced by an electrostatic potential $V(z,x)$ uniform along the $y$-direction and varying between $V_1$ and $V_2$ in the $x$-direction between the two gates, Fig.~\ref{fig:varg_edges}. The electron Hamiltonian is given by
 \begin{equation}
\label{model}
H=-\frac{1}{2m^*}\left(-i\hbar
\nabla-\frac{e{\bf{A}}}{c}\right)^2+e\mathcal E_x x+ \frac{1}{2}
\sigma_z (g^*\mu_B B+J_0 + J_1 x) ~,
\end{equation}
where $\bf{A}$ is a vector potential of a magnetic field ${\bf B}=\curl \bf{A}$, which is directed along negative $z$,
$B= |B_z|$, $m^{*}$,  $e$ and $g^*$ are electron effective mass, charge, and g-factor, $\vec{\sigma}$ is the Pauli matrix vector, $\sigma_z$ is its $z$-component, $\mu_B$ is the Bohr magneton, and $\mathcal E_x = -\nabla_x \int \Psi^*(z) V(z,x)\Psi (z) dz$ is an electric field in $x$-direction caused by the gradient of the gate-induced potential $V(z,x)$. In the mean field approximation, s-d exchange interactions are represented by a uniform part $J_0$ and a gate-induced variation of the s-d exchange $J_1 x$ \cite{sup1}. $J_1$ constitutes a spin-dependent electric field  in $x$-direction. As was demonstrated in \cite{Kazakov2016}, using a combination of front and back gates and in conditions of a non-uniform doping of the quantum well by Mn$^{2+}$ ions along the growth direction $z$, it is possible to achieve almost uniform 2D electron density but induce significant $J_1\gg e\mathcal E_x$, \cite{sup1}. While considering nonzero $\mathcal E_x$ will not change our essential results, we will keep only $J_1$ effective spin electric field and take $\mathcal E_x=0$.

In this model, the electron eigenvalues and wavefunctions are
\begin{eqnarray}
\label{eq:eigen_e_w}
E_{n,s,k_y}&=&\hbar\omega_c\left(n+\frac{1}{2}\right)+\hbar k_y v_{s}-\frac{m^*v_{s}^2}{2}-\frac{s}{2}(g^*\mu_B B+J_0)\\ \label{eq:eigen_w} \psi_{n,s,k_y}&=&u_n\left(x-k_y\ell^2+\frac{v_s}{\omega_c}\right) e^{ik_y y}\chi_s~, \end{eqnarray}
where $\omega_c=eB/(m^*c)$ is the cyclotron frequency, $k_y$ is the $y$-component of the wavevector $\vec{k}$, $u_n$ are the Landau wavefunctions, $s=\pm1$ is for spins up and down, $\chi_{1}=\chi_{\uparrow}=(1,0)^{T}$ and $\chi_{-1}=\chi_{\downarrow}=(0,1)^{T}$ . The spin-dependent  drift velocity $v_s=s\cdot v$, where $v=cJ_1 /2eB$. At $\nu=2$ the edge-like states, Eq.~\ref{eq:eigen_e_w}, are localized near the spectral crossing of $(n=0,\uparrow)$ and $(n=1,\downarrow)$ states and can propagate between the two gated regions with opposite velocities.

A non-magnetic disorder cannot cause scattering between two edge-like states (\ref{eq:eigen_w}) due to their opposite spins. However, two edges with opposite velocities originating from neighboring Landau levels are coupled by spin-orbit interactions, similar to the coupling of edges in a 2D topological insulator introduced by an in-plane Zeeman field. The specific mechanism of such coupling is Rashba (but not the Dresselhaus) spin-orbit interactions, described by a 2D Hamiltonian $H_R= \gamma_R \mathcal E_z (\vec{k}\times \vec{\sigma})_z$. Here ${\mathcal{ E}}_z$ is the component of the electric field perpendicular to the 2D plane, and $\gamma_R$ is the Rashba coefficient. The resulting spin-orbit coupling $h_R= \int \psi^*_{0,1,k_y}H_R\psi_{1,-1,k_y}dxdy$ is given by
\begin{equation}
\label{eq:h_R} h_R=
\sqrt{2}\frac{\gamma_R{\mathcal{ E}}_z}{\ell} e^{-\frac{m^2
\ell^2}{\hbar^2} v^2} \left[1-\frac{m^2
\ell^2}{\hbar^2}v^2\right]~,
\end{equation}
where $\ell=(eB/\hbar c)^{-1/2}$ is the magnetic length.

\begin{figure}
\vspace{-1cm}
\includegraphics[width=0.7\textwidth]{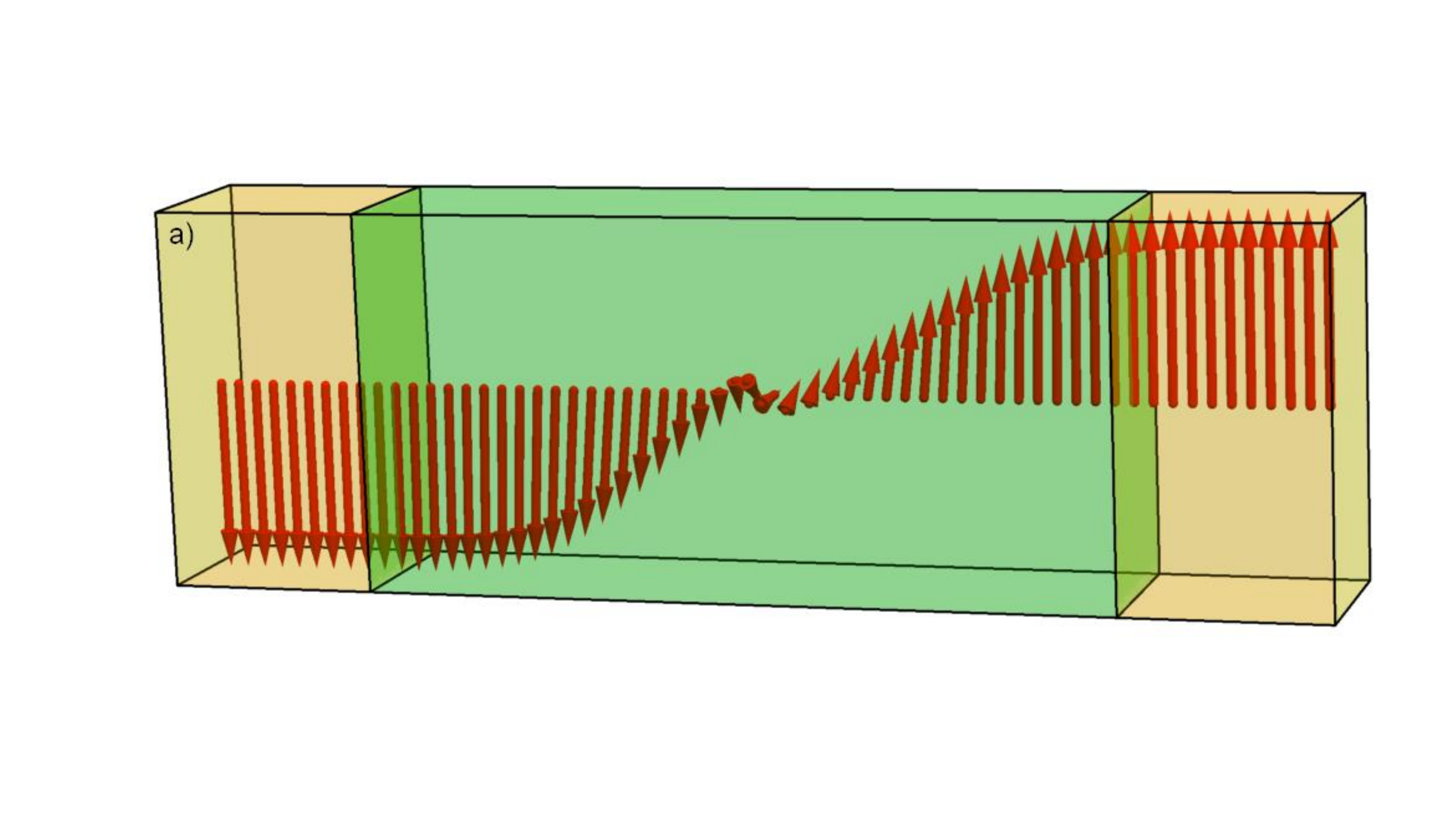}
\includegraphics[width=0.7\textwidth]{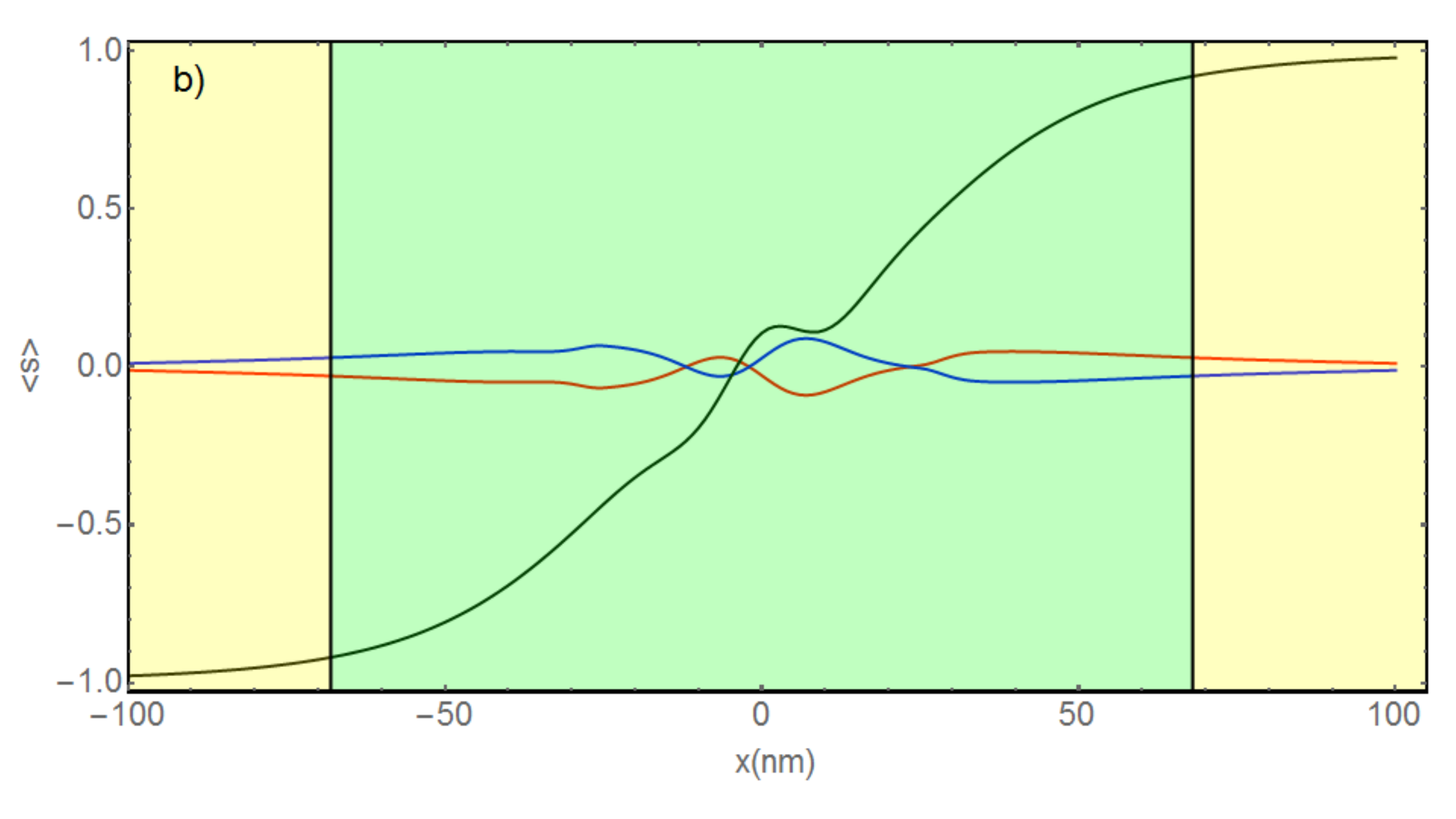}
\caption{Panel a) Spin texture as seen by the ground state in a system of two induced edge states originating from LL1 spin down and LL0 spin up states and coupled by Rashba spin-orbit interactions near spectral crossing. Exchange interactions of electrons are taken into account. Gated areas are shown in yellow, while the
edge channels are propagating through the green region. Panel b)- Average spin projection on
$x$-direction (blue), $y-direction$ (red) and $z$-direction (black) directions in the ground state.}\label{fig:spin_struct}
\end{figure}

In the presence of this spin-orbit coupling, the effective single-particle Hamiltonian in the basis of the  $(n=0, \uparrow)$ and $(n=1, \downarrow)$ states (\ref{eq:eigen_e_w}) near their spectral crossing is given by
\begin{equation}
\label{eq:speffective}
H_e=hk_yv\sigma_z-h_R\sigma_x~.
\end{equation}
Thus, this single-particle system, which serves as a setting for the proximity-induced topological superconductivity, is rather unusual: in contrast to the nanowires and topological insulators, where spin-orbit interactions result in the level crossing and the Zeeman interaction provides a gap, here the Zeeman interaction is responsible for the crossing at $k=0$ while spin-orbit interactions open a gap in the spectrum. The states (\ref{eq:speffective}) exhibit helical electron spin texture similar to the N\'eel domain walls. We have calculated the texture numerically, Fig.\ref{fig:spin_struct}, taking into account exchange interactions between electrons.

In order to see how  non-Abelian quasiparticles can emerge in CdMnTe quantum Hall system, we consider superconductor proximity-induced electron pairing. To illustrate the potential of this system for hosting Majorana modes, we will first assume that the Fermi level is outside the spin-orbit gap and crosses edge-like states forming the helical domain wall. We then consider a proximity effect induced by superconducting Ohmic contacts directly coupling edge-like states to an s-wave superconductor and inducing an order parameter $\Delta(x,y)$. Pairing the states of Hamilonian (\ref{eq:speffective}) is described by the projected order parameter $\Delta_k= \int dxdy \psi_{0,\uparrow,k_y} \Delta(x)  \psi_{1,\downarrow,k_y}$. Due to the opposite velocities of the coupled edge-like states, the $\Delta_k$ is sizable even in the approximation of a constant $\Delta(x,y)=\Delta$  despite different Landau indices for the two edges:
\begin{equation}
\label{Delta_k}
\Delta_k=\Delta e^{ -\frac{m^2 \ell^2} {2\hbar^2}v^2 } \frac{\sqrt{2}m \ell}{\hbar }v.
\end{equation}
The corresponding Bogoliubov-de Gennes ( BdG) equation ${\cal H }\boldsymbol{ \psi} (x,y)=E \boldsymbol {\psi} (x,y)$, where $\boldsymbol {\psi} (x,y)= (u_{\uparrow},u_{\downarrow},v_{\downarrow},-v_{\uparrow})^T$, is defined by
\begin{equation}
\label{eq:bdG_gen}{\cal H }= \left[
\begin{array}{cccc}
\hbar kv -\mu-E_k& -h_R
&  \Delta_k & 0\\
-h_R &- \hbar kv
 -\mu   -E_k & 0 & \Delta_k \\
\Delta_k^* & 0 & -\hbar kv   +\mu -E_k &
-h_R\\
0 &    \Delta_k^* &   -h_R& \hbar kv+\mu -E_k
\end{array}
\right]
\end{equation}
Its four eigenvalues are:
\begin{equation}
E_k=\pm \sqrt{\Delta_k^2+\mu^2+
\hbar^2 k^2v^2 \pm 2\sqrt{\Delta_k^2
h_R^2+\mu^2 h_R^2 +\hbar^2 k^2v^2}}~,
\end{equation}
where $\mu$ is the chemical potential measured from the crossing point energy in the absence of Rashba coupling.  The system becomes gapless for $k=0$ and $\Delta_{k=0}^2+\mu^2=h_R^2$, and at $|h_R|<\sqrt{\Delta_{k=0}^2+\mu^2}$, exhibits a topologically non-trivial superconducting phase. Formally, the emergence  of a topological superconducting phase is somewhat similar to the case of a topological insulator in proximity to an s-wave superconductor \cite{Alicea2012}, but because it is Zeeman splitting that gives level crossing and spin-orbit interactions that leads to the gap here, restriction on the topological phase is defined by the value of the spin-orbit coupling rather than by the Zeeman splitting. It is important to notice that for the chemical potential outside the superconducting gap, i.e., $\mu>h_R$, the induced superconducting order is always topological. Furthermore, topological superconductivity exists even in the absence of a spin-orbit coupling at $h_R=0$. 
Majorana fermions are localized at the contacts between an s-superconductor and a domain wall area.
This Majorana system can be affected by non-magnetic disorder: in contrast to chiral states (\ref{eq:eigen_w}), eigenstates of Hamiltonian (\ref{eq:speffective}) in the presence of the spin-orbit coupling are subject to backscattering similarly to edge states in topological insulators in the presence of Zeeman spin splitting. Backscattering must lead to reduction of domain wall conductance compared to conductance of domain walls formed by chiral states (\ref{eq:eigen_w}), as supported by experimental data on resistance on the flanks of the quantum Hall $\nu=2$ plateau in experiments \cite{Kazakov2017}. 
Thus, for Majorana modes emerging in helical domain wall with the Fermi level positioned outside the spin-orbit gap in the domain wall but inside the quantum Hall gap in the adjacent 2D regions, impurity scattering becomes detrimental  in much the same way as for chiral states in semiconducting wires. Majorana fermions are expected to arise only in a very high mobility quantum Hall samples with small impurity scattering. However, even in this case, due to rather narrow interval of energies, 2D regions show finite conduction at the lowest temperatures, complicating the Majorana setting. 

For chemical potential $\mu$ inside the spin-orbit gap, there exists a signigicant distiction between the present setting and Majorana modes in topological insulators in the presence of Zeeman splitting. In topological insulators, a superconductuctor is often assumed to cover the whole area above the edge states at the sample boundary as opposed to a small contact at the side of the domain wall envisioned here. Correspondingly, certain proximity pairing effect exists throughout topological insulator when $\mu$ is inside the gap, which is characterized by a trivial superconducting phase. In the present setting only very small area defined by a penetration of the wavefunction into an insulating gapped domain wall near the contact can bear some trace of superconductivity, while the rest of the domain wall is generally an insulator. However, as shall see, impurities drastically change this situation.
 
\section{Topological superconductivity generated by disorder}

In order to obtain a well-controlled Majorana setting,  the electron transport has to be conducted exclusively along the helical domain wall. To achieve this, the quantum Hall ferromagnetic transition should be tuned very close to $\nu=2$, where the bulk 2D conduction vanishes. In this case $\mu$ lays inside the spin-orbit gap and conduction is exponentially suppressed at low temperatures in wide regions. However in short helical domain wall channels conduction remains finite, and it was concluded that the in-gap impurity states provide the conduction path \cite{Kazakov2017}. We now show that in the presence of superconducting proximity effect, the helical domain walls with in-gap states can be mapped into a generalized disordered Kitaev chain \cite{kitaev_unpaired_2001,dasarma_chain_2012} where a topologically non-trivial superconducting order and Majorana bound states emerge. 

To consider a superconducting proximity effect in the helical domain walls with the Fermi level  inside the spin-orbit gap in the spectrum of edge states, we first solve a general quantum-mechanical problem  of  impurity-induced states in a magnetic field in the presence of spin-orbit interactions. We then find impurity states in the domain wall in the presence of the mean field gradient of exchange interactions between electrons and Mn ions $J_1$.

\subsection{Effect of spin-orbit coupling on Landau level impurity states}
 
Our goal here is to get analytic results for the impurity-induced states. We model potential variations from remote ionized impurities as short-range potentials with a bound state energy $E_b$ at zero magnetic field. We then solve the system, in which impurity potential is added to Hamiltonian Eq.(\ref{model}).

Short-range impurities in quantizing magnetic field were considered in  \cite{azbel_impurities_1_1993, azbel_impurities_2_1994}. It is convenient to use the wavefunctions of an unbounded electron in a symmetric gauge in a uniform magnetic field,
\begin{eqnarray}
\label{eq:wf_gen}\psi_{n,m,s}^0(r,\varphi)&=&\sqrt{\frac{n!}{(n+m)!2^{m+1}\pi
\ell}} e^{\left( i m \varphi+ \frac{i}{4} \frac{r^2}{\ell ^2}
\sin(2\varphi)-\frac{r^2}{4\ell ^2}\right)} \left(\frac{r}{\ell}\right)^m
L_{n}^m\left(\frac{r^2}{2 \ell^2}\right)\chi_{s}~,
\end{eqnarray}
corresponding to degenerate states with energy $E_{n,m,s}^0=\hbar\omega_c \left(n+\frac{1}{2}\right)+s V_z$, $s=\pm 1$, $V_z$ is the spin splitting that includes band Zeeman effect and mean field exchange splitting due to the electron spin interaction with Mn spins, $L_n^m$ denotes
the Laguerre polynomials, $r$ and  $\varphi$ are the polar coordinates, $n\geq 0$ and
$m\geq-n$ are integers, and
$\chi_1=\chi_{\uparrow}=(1,0)^{T}$ and $\chi_{-1}=\chi_{\downarrow}=(0,1)^{T}$ are the spinors.

Following \cite{azbel_impurities_1_1993, azbel_impurities_2_1994} we begin with considering a single impurity  at the origin in the presence of the Landau quantization.
The short-range impurity does not affect states with $m\neq0$ as their wavefunction
is zero in the origin, and all states with $m\ne 0$ are still described by the wavefunctions given by Eq. (\ref{eq:wf_gen}) and the corresponding eigenenergies $E_{n,m\ne 0,s}^0$. The states with $m=0$ are bounded by the
impurity and the energy and wavefunctions of these states are:
\begin{eqnarray}
\label{eq:en_bound}E_{n,0,s}^0&=&\hbar \omega_c\left(n+\frac{1}{2}-\delta_n\right)+s V_z\\
\psi_{n,0,s}^0&=&\frac{|\Gamma(-n+\delta_n)|}{\sqrt{\pi
\Psi'(-n+\delta_n)}}\frac{(-1)^n}{r}
\label{eq:wf_bound}W_{n+\frac{1}{2}-\delta_n,0}\left(\frac{r^2}{2}
\right)\chi_s~,
\end{eqnarray}
where $W$ is the Whittaker function and $\Psi$ is the digamma function.
In a high magnetic field limit the impurity split-off $\delta_n$ is given by
\begin{equation}
\label{eq:deltan} \delta_n= \left|\Psi(n+1)-\ln\frac{|E_b|}
{\hbar\omega_c}\right|^{-1}~,
\end{equation}
For states with $\delta_n \ll1$ the digamma
function in Eq. (\ref{eq:deltan}) is much smaller than the logarithmic
part and $\delta_n=1/\ln( \hbar\omega_c/|E_b|)\equiv \delta$ is independent of
$n$. To simplify our analysis, we will consider this approximation; our conclusions, however, are quite general and this
restriction is not crucial.

We now include the Rashba Hamiltonian $H_R$ using the basis set that includes
the orthonormalized wavefunctions Eq. (\ref{eq:wf_gen}) for $m\neq 0$ and wavefunction determined by Eq. (\ref{eq:wf_bound}) for
$m=0$. The non-zero matrix elements (neglecting terms of order of  $\mathcal{O}(\beta
\delta/\hbar\omega_c )$ and $\mathcal{O} (\delta^2$) ) are
\begin{equation}
\bra{\psi_{n,m-1,\uparrow}}H_R\ket{\psi_{n-1,m,\downarrow} } = \frac{\beta \sqrt{2
n}} {\ell} =\Delta_{so}\sqrt{n} ~,
\end{equation}
where $\beta=\gamma_R\mathcal{ E}_z$. $\Delta_{so}$ coinsides with $h_R$  given by Eq. (\ref{eq:h_R}) at $n=1$  when $J_1$ is neglected, i.e., at $v=0$.

\begin{figure}
\includegraphics[width=0.95\textwidth]{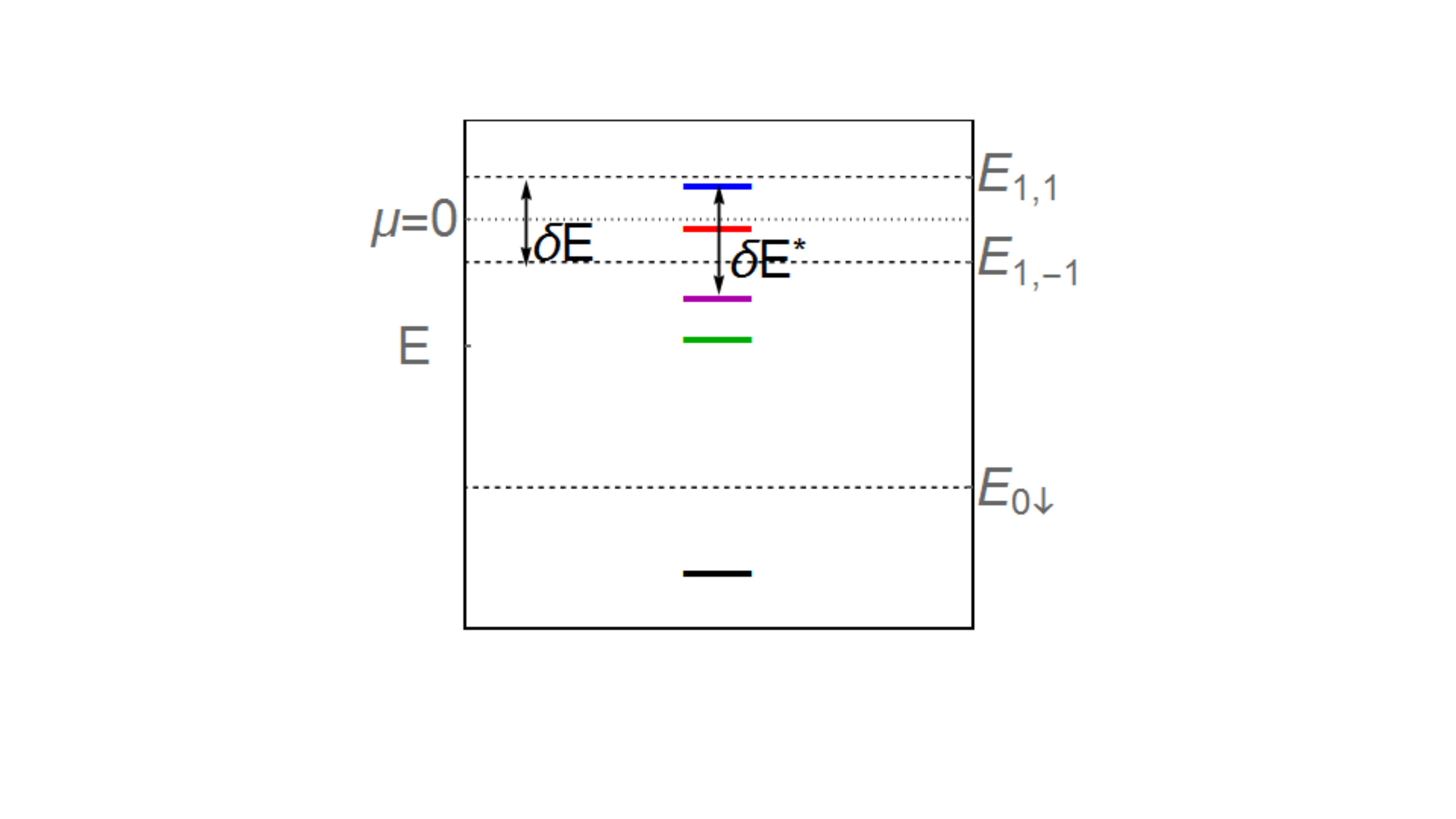}
\caption{Electron energy spectrum in quantized magnetic field in the presence of attractive impurity center and spin-orbit interactions. Splitting of the degenerate in $m$ $(n=1, s=1)$ and $(n=1, s=-1)$  levels  is caused by the cyclotron splitting and spin splitting due to Zeemann, exchange and spin-orbit interactions.  Each impurity results in two energy levels given by Eqs. (\ref{eq:e_so_bound_1_bulk},\ref{eq:e_so_bound_2_bulk}) due to two linear combinations of m=0 and m=-1 states for each of the spin-resolved Landau levels. Impurity-bound state $E_{1,1}^{+}$ is shown in blue, $E_{1,1}^{-}$ is shown in red, 
$E_{1,-1}^{+}$ is shown in magenta and $E_{1,-1}^{-}$ is shown in green.
 Only one impurity induced state is present for $(n=0, \downarrow)$ Landau level, which is not affected by spin-orbit coupling, shown in black.}
\label{FigSpectrumcross-section}
\end{figure}

The effect of spin-orbit interaction on Landau electron states and impurity bound states in quantized magnetic field is two-fold.
First, for all states except the lowest $n=0$ Landau level with spin down, spin-orbit interaction leads to an additional repulsion of Landau states $(n, [m\ne -1,0], \uparrow)$ and $(n+1, [m\ne -1,0],\downarrow)$  and results in energy series
\begin{equation}
\label{free}
E_{n,m,s}=\hbar \omega_c n +s \sqrt{\frac{ 2 n\beta^2}{ \ell^2} +\left( \frac{1}{2}\hbar\omega_c-V_z\right)^2}~,~m\neq -1,0~,
\end{equation}
where $n\ge 1$, and $s=\pm1$ describes spin states. The nondegenerate state with $n=0$ has energy $E_0= \hbar\omega_c/2 -V_z$, and is a ground state for $\Delta_{so}\ll \hbar\omega_c$ considered here. In Eq.(\ref{free}), for a pair of states at a given n, $s=1$ characterises the electron state with bigger energy, while state with lower energy is characterized by $s=-1$. State $(n,s=1)$ originates from state $(n,\downarrow)$ in the absence of spin-orbit coupling, while the state $(n, s=-1)$ originates from the state $(n-1,\uparrow)$. Except for the exclusion of states with $m=0$ and $m=-1$ this is the Rashba spectrum for conduction electrons \cite{rashba60}. 
Energy separation 
$\delta E= E_{1,m,+1}- E_{1,m,-1}$ arising from cyclotron splitting as well as spin splitting due to Zeemann, exchange and spin-orbit interactions  is given by
\begin{equation}
\delta E= 2 \sqrt{\frac{ 2\beta^2}{ \ell^2} +\left( \frac{1}{2}\hbar\omega_c-V_z\right)^2}.
\end{equation}
At
Zeeman energy $V_z=g^*\mu_B B+J_0=\hbar \omega_c/2$, 
energy states $(n, s=1)$ and $(n, s=-1)$, and particularly $(n=1, s=1)$ and $(n=1,s=-1)$ energy states are degenerate in the absence of spin-orbit interactions, but are splitted in its presence, with energies
$E_{\pm}=\hbar\omega_c\pm \Delta_{so}$.
Second, spin-orbit interaction couples $(n-1,  m=0, \uparrow)$ impurity-bound state with  $(n,  m=-1, \downarrow)$ Landau level state, as well as
$(n,  m=0, \downarrow)$ impurity-bound state with  $(n-1,  m=-1,\uparrow)$ Landau level state. Such coupling introduces level repulsion within these pairs of coupled states, which results in the splitting of the $m=-1$ levels off the angular-momentum--degenerate Landau levels. Therefore, we now have two bound states for each spin-resolved Landau level, defined by two linear combinations of $m=0$ and $m=1$ states; the exception is a single bound state associated with the lowest $(n=0,\downarrow)$ Landau level. Energies of the two series of bound states at $n>0$ are given by:
\begin{eqnarray}
\label{eq:e_so_bound_1_bulk} E_{n,\varsigma}^{+}&=&\hbar \omega_c
\left( n- \frac{\delta}{2} \right)+
\varsigma \sqrt{\left(\hbar \omega_c  \frac{1-\delta }{2}-V_z\right)^2+\frac{ 2 n\beta^2}{ \ell^2}} ~, \\
\label{eq:e_so_bound_2_bulk} E_{n,\varsigma}^{-}&=&\hbar \omega_c
\left( n- \frac{\delta}{2} \right)+
\varsigma \sqrt{\left(\hbar \omega_c  \frac{1+\delta}{2}-V_z\right)^2+\frac{ 2 n\beta^2}{ \ell^2}} ~,
\end{eqnarray}
where $\varsigma=\pm 1$ denotes two different superpositions of $m=0$ and $m=-1$ states for a given $n$ in each of the series. Impurity-bound states $E_{n,\varsigma}^{+}$ originate from $(n,  m=-1, \downarrow)$ Landau level states, and   states $E_{n,\varsigma}^{-}$ originate from $(n-1,  m=-1, \downarrow)$ Landau level states.  The electron and impurity-bound energy levels in quantized magnetic field in a quantum well in the presence of Rashba interactions are shown in Fig. \ref{FigSpectrumcross-section}. As follows from Eqs. (\ref{eq:e_so_bound_1_bulk},\ref{eq:e_so_bound_2_bulk}), splitting of levels with opposite $\varsigma$ in the same series, e.g., $\delta E^*$ in Fig.\ref{FigSpectrumcross-section} is bigger than splitting $\delta E$ between coupled Landau levels due to additional level repulsion caused by impurity split-off $\hbar\omega_c\delta$.   

\begin{figure}
\includegraphics[width=0.95\textwidth]{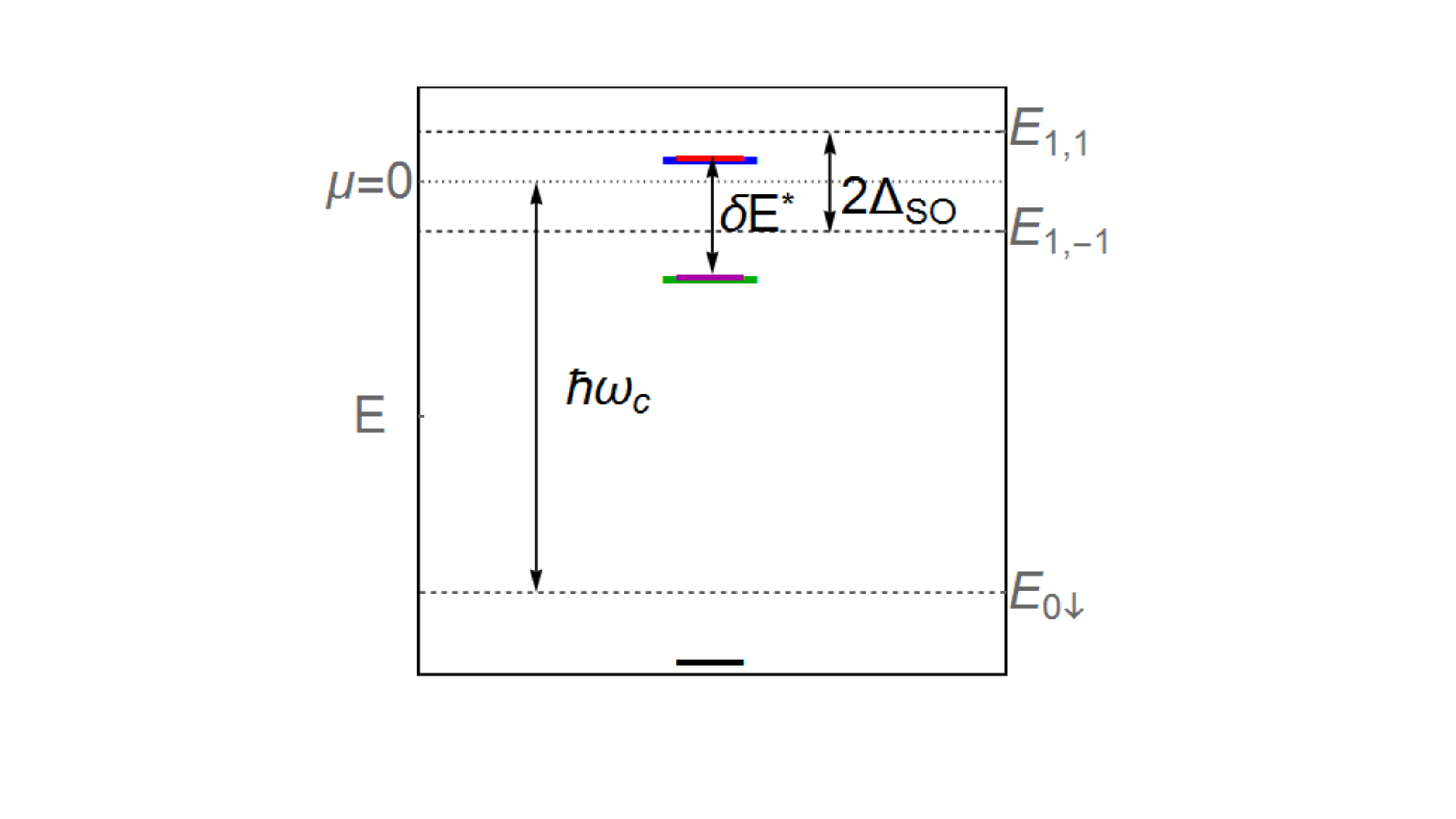}
\caption{Electron energy spectrum in quantized magnetic field in the presence of attractive impurity center and spin-orbit interactions,  in the case of compensation between cyclotron splitting and the sum of the Zeemann and exchange interactions  leading to degeneracy of $(n=0, \uparrow)$ and $(n=1, \downarrow)$ Landau levels given by Eq.(10). The splitting $2\Delta_{so}$ of the $(n=1, s=1)$ and $(n=1, s=-1)$ unbound states is due to Rashba coupling only.
Impurity levels from series + and -  in Eqs. (\ref{eq:e_so_bound_1_bulk},\ref{eq:e_so_bound_2_bulk})  become degenerate,
 so that $E_{1,1}^{+}=E_{1,1}^{-}$  and $E_{1,-1}^{+}=E_{1,-1}^{-}$ (shown as coincidence of blue and red and coincidence of magenta and green). Splitting between pairs of degenerate levels  $\delta E^*$ is due to both Rashba coupling and level repulsion caused by impurity split-off $\hbar\omega_c\delta$.}
\label{FigSpectrumCompensation}
\end{figure}

At  Zeeman energy $V^*_z=g^*\mu_B B+J_0=\hbar \omega_c/2$, levels of series $ E_{n,\varsigma}^{+}$ and $ E_{n,\varsigma}^{-}$ become degenerate. In particular,
the double degenerate level
\begin{equation}
\label{main}
E^{*}_{n=1, +}=\hbar\omega_c\left(1-\frac{\delta}{2}\right)+\frac{1}{2} \sqrt{4h_R^2+(\hbar\omega_c\delta)^2}
\end{equation}
corresponding to $\varsigma= 1$  lies in between $(n=1,s=1)$ and $(n=1,s=-1)$  levels, and a double degenerate level at $\varsigma=- 1$ with
energy 
\begin{equation}
 E^{*}_{n=1, -}=\hbar\omega_c(1-\frac{\delta}{2})-\frac{1}{2} \sqrt{4h_R^2+(\hbar\omega_c\delta)^2}
\end{equation}
 lies below
$(n=1,s=-1)$ level. Remarkably, degenerate impurity-bound states with energy  $E^{*}_{n, +}$ have opposite spins, and  states $E^{*}_{n=1, -}$ also have opposite spins. This is a consequence of degeneracy between $(n-1,\uparrow)$ and $(n,\downarrow)$  Landau levels when spin-orbit interactions are not included.

Wavefunctions of degenerate Landau levels and impurity split-off states in the bulk in the presence of spin-orbit interactions for arbitrary $n>0$ and $V_z=V^*_z$ can be written as:
\begin{eqnarray}
\label{eq:wf_so_gen}
\psi_{n,m,\varsigma}&=&\frac{1}{\sqrt{2}}\left(
\psi^0_{n,m,1}-\varsigma \psi^0_{n-1,m+1,-1} \right) ~,~m\neq -1,0~,\\
\label{eq:wf_so_bound} \psi^1_{n,m,\varsigma}&=&\frac{1}{\sqrt{2}}
\left( \varrho_{\beta,\delta}^{(-1)^m\varsigma,n}
\psi^0_{n,m,1}-\varsigma
\varrho_{\beta,\delta}^{(-1)^{m+1}\varsigma,n} \psi^0_{n-1,m+1,-1}
\right) ~,~m=-1,0~,
\end{eqnarray}
where wavefunctions $\psi^0_{n,m,s}$ are defined by Eq.(\ref{eq:wf_gen}), and
\begin{equation}
\varrho_{\beta,\delta}^{\pm 1,n}= \sqrt{1\pm\frac{ \hbar \omega_c
\ell \delta} {\sqrt{\left( \hbar \omega_c \ell \delta\right)^2+ 8
n\beta^2 }}}.
\end{equation}
 For $n=0$, the eigenstates are defined by Eqs. (10, 11): $\psi_{0,m,1}^1=\psi_{0,m,1}^0$ and
$E_{0,m,1}^1=E_{0,m,1}^0$.

\subsection{Impurity states in a helical domain wall}

\begin{figure}
\includegraphics[width=0.9\textwidth]{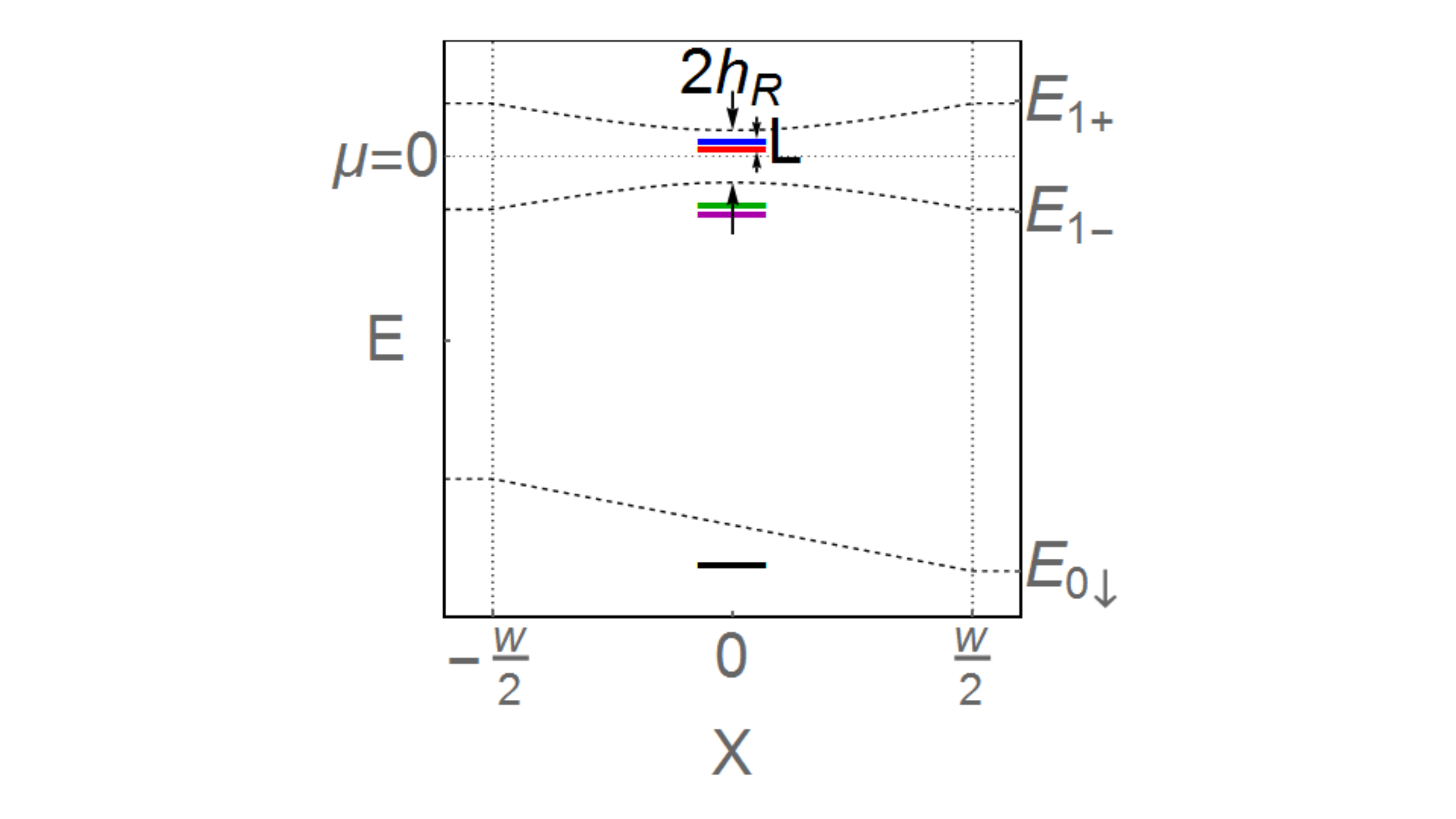}
\caption{Electron spectrum in the presence of impurities in a helical domain wall of width $W$.  In the presence of compensation  between cyclotron energy and the sum of Zeemann and exchange energies for electrons in the helical domain wall, splitting between red-blue and green-magenta doublets of impurity levels in the center of the channel is due to spin-orbit interactions. Electron edge states there are also separated only by the spin-orbit coupling. 
Red and blue levels (and green and magenta levels) are separated in energy due to angular momentum splitting arising because of the effective spin-dependent electric field $J_1$. 
One impurity doublet (red-blue) falls into the spin-orbit gap  between edge states arising from $(n=1, s=1)$ and  $(n=1, s=-1)$ 2D Landau states (dashed lines). The other doublet (green-magenta) is below the spin-orbit gap. Only a single non-degenerate level (black solid segment) is split off $(n=0, \downarrow)$  Landau level shown by the lower black dashed line.}
\label{FigChannellSpectrum}
\end{figure}

So far we discussed the bulk Landau levels and the bulk impurity states in the presence of spin-orbit coupling. In the presence of  the spin-dependent electric field $J_1$  in a narrow range of coordinate $x$, which leads to the formation of a helical domain wall, these bulk states change in a two-fold way. First, Landau levels with multiple degeneracy in angular momenta (\ref{eq:wf_so_gen}) form  linear combinations that correspond to an edge-like states (\ref{eq:eigen_e_w}), which are gapped by spin-orbit interactions and described by the effective Hamiltonian (\ref{eq:speffective}). Two doublets of impurity states also evolve,  Fig.\ref{FigChannellSpectrum}: one doublet with $\varsigma=+1$  falls into the gap between spin-orbit split edge-like states, and the other doublet with $\varsigma=-1$ is below the spin-orbit gap. The second effect of the effective spin-dependent electric field $J_1$ is angular momentum splitting of the in-gap impurity states. The angular momentum splitting of the $E^{*}_{1, +}$ double degenerate level (\ref{main}) for an impurity centered at the origin in the area of the helical domain wall, is given by
\begin{equation}
\label{eq:en_so_v_bound} L= \frac{
2\hbar^2 v^2}{ \ell
\sqrt{ \left(\hbar \omega_c \delta \ell \right)^2+8\beta^2}
}.
\end{equation}
Angular momentum splitting $L$ arises in the second order in the effective spin-dependent electric field $J_1$, and therefore is quadratic in $v$.

\subsection{Chain of impurity states}

\begin{figure}
\includegraphics[width=0.8\textwidth]{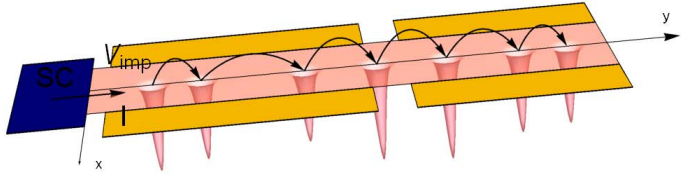}
\vspace{-3cm}
\caption{Schematic view of the conducting channel with
proximity induced superconductivity (blue contact), with attractive impurity potential (red)}
\label{FigChannell}
\end{figure}

Our goal is to study a chain of in-gap states,  Fig. \ref{FigChannell}. For impurity potentials centered at ${\bf{R}}_k=(X_k,Y_k)$, their separation along the $y$-direction is assumed much larger than the width of a helical domain wall. 
Therefore the chain can be considered as one-dimensional, with ${\bf{R}}_k=(X_k=0,Y_k)$. Also, in high magnetic field
$|R_k-R_{k-1}|\gg \ell$.
We will assume that impurity centers may have slightly different binding energies and therefore different impurity split-offs $\delta$, e.g.,  because of their varying $z$-coordinate in a doping layer and therefore varying separation from the quantum well. We will denote the split-off for an impurity centered at ${\bf{R}}_k$ as $\delta_k$.  Angular momentun splittings $L_k$ for impurity sites centered at ${\bf{R}}_k$ also differ from site to site:
\begin{equation}
L_k= \frac{2\hbar^2 v^2}{ \ell \sqrt{ \left(\hbar \omega_c
\delta_k \ell\right)^2+8 \beta^2} }.
\end{equation}
  The wavefunctions of electrons bound to a single impurity are given by
\begin{equation}
\label{eq:wavechain}
\psi_m^{(k)}({\bf{r}})=\psi_{1,m,-1}
({\bf{r}}-{\bf{R}}_k)e^{iX_ky/\ell^2}=\psi_{1,m,-1}({\bf{r}}-{\bf{R}}_k).
\end{equation}
Considering a chain, we orthogonolize these wavefunctions assuming that only overlap between wavefunctions of electrons centered on the nearest neighbors is essential. The orthonormalized wavefunctions are:
\begin{eqnarray}
\ket{\tilde \psi_{m}^{(k)}}&&\simeq\ket{\psi_m^{(k)}} -
\frac{1}{2}\sum_{m_1=-1}^0\ket{\psi_{m_1}^{(k+1)}}
S^{k+1,k}_{m_1,m}\nonumber\\
&&-\frac{1}{2}\sum_{m_2=-1}^0\ket{\psi_{m_2}^{(k-1)}}
S^{k-1,k}_{m_2,m}~,\hspace{3mm} m=-1,0~,
\end{eqnarray}
where the overlap integrals of the electron wavefunctions on isolated centers located at  $\bf{R}_k^{\prime}$ and $ \bf{R}_k $ are given by
\begin{equation}
S^{k^{\prime},k}_{m^{\prime},m}=\bra{\psi_{m^{\prime}}^{k^{\prime}}}\ket{\psi_{m}^{(k)}}.
\end{equation}
We seek the wavefunctions of the Hamiltonian of the chain
\begin{equation}
H=-\frac{1}{2m^*}\left(-i\hbar
\nabla-\frac{e{\bf{A}}}{c}\right)^2+V_z \sigma_z + \sum_k U( {\bf r},\bf{R}_k)
\end{equation}
in the form
\begin{equation}
\Psi= \sum_{m,k} a_{mk} \ket{\tilde \psi_{m}^{(k)}}~,\hspace{3mm} m=-1,0~.
\end{equation}
Then the effective Hamiltonian $H_{mk,m^{\prime}k^{\prime}}$ acting on coefficients  $a_{mk}$ is defined by remormalized single-impurity site energies $\tilde{E_{mk}} = \bra{\tilde\psi_{m}^{k}} H \ket{\tilde\psi_{m}^{k}}$ and tunneling matrix elements
$w^{m^{\prime},m}_{k^{\prime},k}=\bra{\tilde\psi_{m^{\prime}}^{k^{\prime}}} H \ket{\tilde\psi_{m}^{k}}$. The leading contribution to tunnelling arises from matrix elements
\begin{eqnarray}
\label{eq:mat_el_tun_s} w_{k+1,k}^{m,m} &\simeq&
\tilde \delta_{k+1,k}  (-1)^{m+1} P_{k+1,k}^
{\beta} \frac{1}{4} \left(\frac{Y_{k+1,k}}{\sqrt{2}\ell}\right)^{2m+2} e^{-\frac{Y_{k,k+1}^2} {4\ell^2}}~,\\
\label{eq:mat_el_tun_o} w_{k+1,k}^{-1-m,m} &\simeq&
 \delta^{d}_{k+1,k} Q_{k+1,k}^{\beta} \frac{1}{4}
\frac{Y_{k+1,k}}{\sqrt{2}\ell} e^{-\frac{Y_{k,k+1}^2}{4\ell^2}}~,
\end{eqnarray}
where
\begin{eqnarray}
P_{k+1,k} ^{\beta}&=& 1-\frac{\hbar \omega_c \ell \tilde
\delta_{k+1,k}}{\sqrt{( \hbar \omega_c \tilde \delta_{k+1,k}
\ell)^2+8 \beta^2}}~,\\
Q_{k+1,k}^{\beta}&=& \frac{\beta}{\sqrt{( \hbar \omega_c \tilde
\delta_{k+1,k} \ell)^2+8 \beta^2}}~,
\end{eqnarray}
$\tilde \delta_{k+1,k}=(\delta_{k}+\delta_{k+1})/2$ is an average split-off of the neighboring impurity centers, $\delta^{d}_{k+1,k}=\delta_{k}-\delta_{k+1} $, and
$Y_{k+1,k}=Y_{k+1}-Y_{k}$ . These expressions are obtained by
expanding the overlap matrix elements and keeping only the leading
terms in $1/Y_{k+1,k}$, $e^{-Y_{k+1,k}^2}$ and $ \delta^{d}$.

\subsection{Superconducting coupling}

We project electron interactions due to the proximity-induced superconducting paring $H_{\Delta}=\Delta \int
{\hat\psi^{\dagger}_{\uparrow} \hat\psi^{\dagger}_{\downarrow}}+
h.c.$, onto the Hilbert space of bound states
$\psi^{(k)}_{m}$. As we are interested here is a single superconducting contact to a quantum Hall system, phase of the order parameter is unimportant and we  take
$\Delta>0$ without the loss of generality. The effective Hamiltonian for the superconducting pairing with a chain of impurity states then reads
\begin{equation}
\label{eq:gen Delta}H_{\Delta}\simeq \sum_{k}\tilde\Delta_k
c_{k,0}^{\dagger} c_{k,-1}^{\dagger}+
\sum_{m,m'=-1,0}\Delta_{k,k+1}^{m,m'} c_{k,m}^{\dagger}
c_{k+1,m'}^{\dagger} + h.c.~,
\end{equation}
where
\begin{eqnarray}
\label{eq:eff_d_single}
\tilde\Delta_k&=&\Delta\frac{1-\gamma_0\delta_k}{\sqrt{8}}\\
\label{eq:eff_d_same} \Delta_{k,k+1}^{m,m}&=& \Delta i
(4m+3)\left(\frac{Y_{k,k+1} }{\sqrt{2} \ell} \right) ^{2m+1}
Q_{k+1,k}^{\beta}e^{-\frac{Y_{k+1,k}^2}{8
\ell^2}}\\
\label{eq:eff_d_diff} \Delta_{k,k+1}^{-1-m,m}&=& \Delta (-1)^m
(4m+3) \left(\frac{Y_{k,k+1}} {\sqrt{2} \ell}\right) ^{2}
\left(P_{k+1,k}^{\beta}+m-\frac{1}{2}\right)e^{-\frac{Y_{k+1,k}^2}{8
\ell^2}}~,
\end{eqnarray}
$\gamma_0\simeq 1.89258$ is a numerical constant, and $m=-1,0$.

\subsection{Single impurity site in the presence of superconducting pairing}

In order to address  the topological superconductivity and Majorana fermions in a chain of impurity states, we first
consider a single site in the presence of superconducting coupling within the Bogoliubov-DeGennes formalism.
We restrict the Hilbert space to $\psi^{1}_{1,0,-1}$ and
$\psi^{1}_{1,-1,-1}$ near impurity site $k$ with
coordinates $R_k$. We denote electron creation operators for these
states $c_{k,+1}^{\dagger}$ and $c_{k,-1}^{\dagger}$.
Then the effective Hamiltonian is given by
\begin{equation}
H_{k}=\sum_{i,j}{(\varepsilon_k+L_k\sigma_z)_{i,j} c_{k,
i}^{\dagger} c_{k, j}+i \tilde\Delta_k \hat c_{k, i}^{\dagger}
\left(\sigma_{y}\right)_{i,j} \hat c_{k, j}^{\dagger}-i \tilde\Delta_k
\hat c_{k, i} \left(\sigma_{y}\right)_{i,j} \hat c_{k, j}}~,
\end{equation}
where $\mu$ is the chemical potential, and on-site energies are
\begin{equation}
\varepsilon_k=-\hbar \omega_{c}  \frac{\delta_{k}}{2} +
\frac{1}{2} \sqrt{ \left(\hbar \omega_{c} \delta_{k} \right)^2+8
\left(\frac{\beta}{\ell}\right)^2}-\mu~,
\end{equation}
where $\tilde \Delta_k$ is defined by
Eq.~(\ref{eq:eff_d_single}). We diagonaize this Hamiltonian using the
Bogoliubov transformation
\begin{equation}
\label{eq:a_bog_pm} \hat a_{k,\pm} =  \pm \sqrt{1+
\frac{\varepsilon_k}{\sqrt{\varepsilon_k^2+|\tilde\Delta_k|^2}}}
e^{i\frac{\pi}{4}} \hat c_{k,\pm 1}+ \sqrt{1-\frac{\varepsilon_k}
{\sqrt{\varepsilon_k^2+|\tilde \Delta_k|^2}}} e^{i\frac{\pi}{4}}
c_{k,\mp 1}^{\dagger}
\end{equation}
that gives eigenvalues $\mu_{k}\pm L_k$, where
\begin{equation}
\mu_{k}=\sqrt{\tilde \Delta_k^2+\epsilon_{k}^2}.
\end{equation}

\subsection{Topological superconductivity in a chain of impurity-bound states}

We now  study a chain of impurity-bound sites placed at
${\bf{R}}_{k}=(0,Y_k)$. We denote ${\bf{R}}_{k,k+1}={\bf{R}}_{k+1}
-{\bf{R}}_k$.
The Hamiltonian of the chain is defined by the single site energies, superconducting coupling and inter-site tunneling:
\begin{equation}
\label{eq:all_chain} H_{c}=\sum_k H_k+\sum_{k,i,j} w_{k+1,k}^{i,j}
\hat c_{k+1,i}^{\dagger} \hat c_{k,j}+\sum_{k,i,j}
\Delta_{k+1,k}^{i,j} c_{k+1,i}^{\dagger} \hat c_{k,j}^{\dagger}
+h.c~,
\end{equation}
where $w_{k+1,k}^{i,j}$ are given by Eqs.
(\ref{eq:mat_el_tun_s}), (\ref{eq:mat_el_tun_o}) and
$\Delta_{k+1,k}^{i,j}$ are given by Eqs.
(\ref{eq:eff_d_same}) and (\ref{eq:eff_d_diff}). Analogous to \cite{Akhmerov_chain_2013}, we project the
Hamiltonian $H_c$  onto the subspace of fermionic excitations given by $a_{k,-}$ on each site. These excitations are defined by  Eq.(\ref{eq:a_bog_pm}). Then the effective Hamiltonian is
\begin{equation}
\label{eq:gen_kitaev} H=\sum_k\left[\left(\mu_{k}-Z_k\right) \hat
a_{k,-}^{\dagger} \hat a_{k,-}+t_k \hat a_{k+1,-}^{\dagger} \hat
a_{k,-}+ \bar\Delta_k \hat a_{k+1,-}^{\dagger} \hat
a_{k,-}^{\dagger}\right]+h.c.~,
\end{equation}
where in the leading approximation
\begin{eqnarray}
t_k&=&\Delta
\frac{\sqrt{2}}{4}\left(\frac{Y_{k+1,k}}{\sqrt{2}\ell}\right)^2
r_{k,\delta}\sqrt{1+r_{k,\delta}^2}\left(P_{k+1,k}^{\beta}-\frac{3}{4}\right) e^{-\frac{Y_{k+1,k}^2}{8 \ell^2}}~,\\
\bar\Delta_k&=&\Delta
\frac{3}{16}\left(\frac{Y_{k+1,k}}{\sqrt{2}\ell}\right)^3
\sqrt{1+r_{k,\delta}^2}
\left(\sqrt{1+r_{k,\delta}^2}-1\right)Q_{k+1,k}^{\beta}e^{-\frac{Y_{k+1,k}^2}{8
\ell^2}} ~,
\end{eqnarray}
$\mu_{k+1,k}=\left(\mu_k+\mu_{k+1} \right)/2$ and
$r_{k,\delta}=\tilde\Delta/\mu_{k+1,k}$.

The term proportional to $\bar\Delta_k$ constitutes a $p$-type superconducting pairing. We therefore arrived at
a generalized version \cite{dasarma_chain_2012,Akhmerov_chain_2013}
of the Kitaev chain \cite{kitaev_unpaired_2001}. Except possibly for the
$\left(P_{k+1,k}^{\beta}-\frac{3}{4}\right)$ factors appearing in
the definition of an effective tunneling amplitudes $t_k$,   $t_k$ and a superconducting pairing
$\bar\Delta_k$ do not change sign from site to site. $t_k$  becomes
zero when $\hbar\omega_c\delta_{k}=\beta/\ell\sqrt{15}$.
However,  $\mu$ can be adjusted so that $t_k>0$ in a chain of impurity sites.

\begin{figure}
\includegraphics[width=0.7\textwidth]{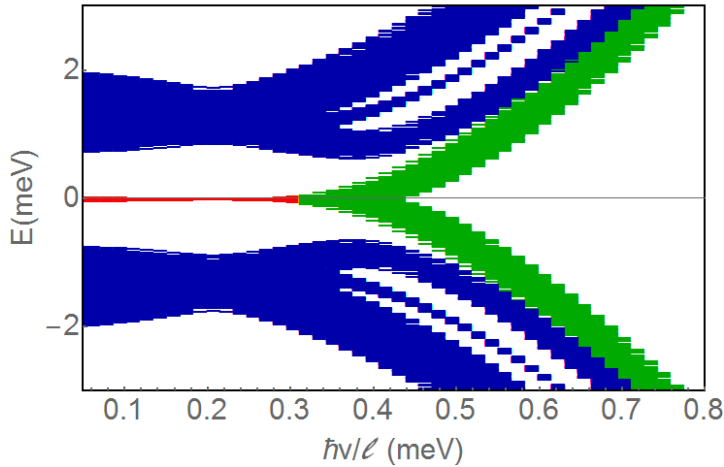}
\caption{Spectra of 100 realizations of a chain of 5 localized
states with superconducting coupling. The total length of the chain
is 200 nm. Bound states energies $\in[\mu-k_BT,\mu+k_BT]$, where
$T=0.1 {\rm{K}}$. Minimal separation of centers of localized states is
25 nm, $\Delta=0.1 {\rm{meV}}$, $\gamma_R=0.44 {\rm{meV nm}}$   $\mu=32{\rm{\mu eV}}$. In-gap
states that disappear with increasing velocity  (transition from red to green) signify
the existence of the Majorana bound states (red).} \label{FigSpectrumw1}
\end{figure}

We thus arrive to the realization of a sign ordered Kitaev
chain \cite{dasarma_chain_2012} that supports
two Majorana localized modes at its ends if
$|\mu_k-L_k|<\max(t_{k+1},\bar\Delta_{k+1})$.  Although this criterion creates an impression that it can possibly be satisfied even at $L_k=0$, it is important to keep in mind that non-zero $L_k>k_BT$  is an important factor that prevents fermion doubling. $L_k$ separates two angular momentum/spin species of in-gap states proximity-coupled to a superconductor. That constitutes a difference between the setting of in-gap Majorana modes and Majorana modes in a topological insulator. In a topological insulator proximity-coupled to a superconductor, Majorana modes can emerge at zero Zeeman splitting because fermion doubling in topological insulator is removed by the chiral character of spin edge states. However, the in-gap electron states in our setting do not propagate, and are not characterized by a wavevector.  In the absence of
 $J_1$ defining the velocity $v$ of the edge states, the states are degenerate in angular momentum and spin simultaneously, which leads to the fermion doubling.
However, the gradient of exchange interactions results in angular momentum splitting $L_k$ that removes fermion doubling, and leads to the emergence of topological superconductivity.
In Fig. \ref{FigSpectrumw1}, we present numerically calculated spectra of a short chain of localized states with proximity-induced superconducting coupling. At small $J_1$ and $v$ ($\hbar v/\ell<0.3$) the chain is characterized by zero modes, but for larger $J_1$ and $v$ ($\hbar v/\ell > 0.3$) states inside the superconducting gap disappear. The condition $\hbar v/\ell \approx 0.3$ corresponds to the topological phase transition.

Tuning the angular momentum splitting, we can bring the system in and out of the topological phase, creating and destroying Majorana modes at the end of the chain. $L_k$, in contrast to settings described in \cite{dasarma_chain_2012, Akhmerov_chain_2013} is unrelated to the value of a magnetic field, but is defined by velocities of gapped edge channels $v$, which are controlled by electrostatic gates.

\begin{figure}
\includegraphics[width=\textwidth]{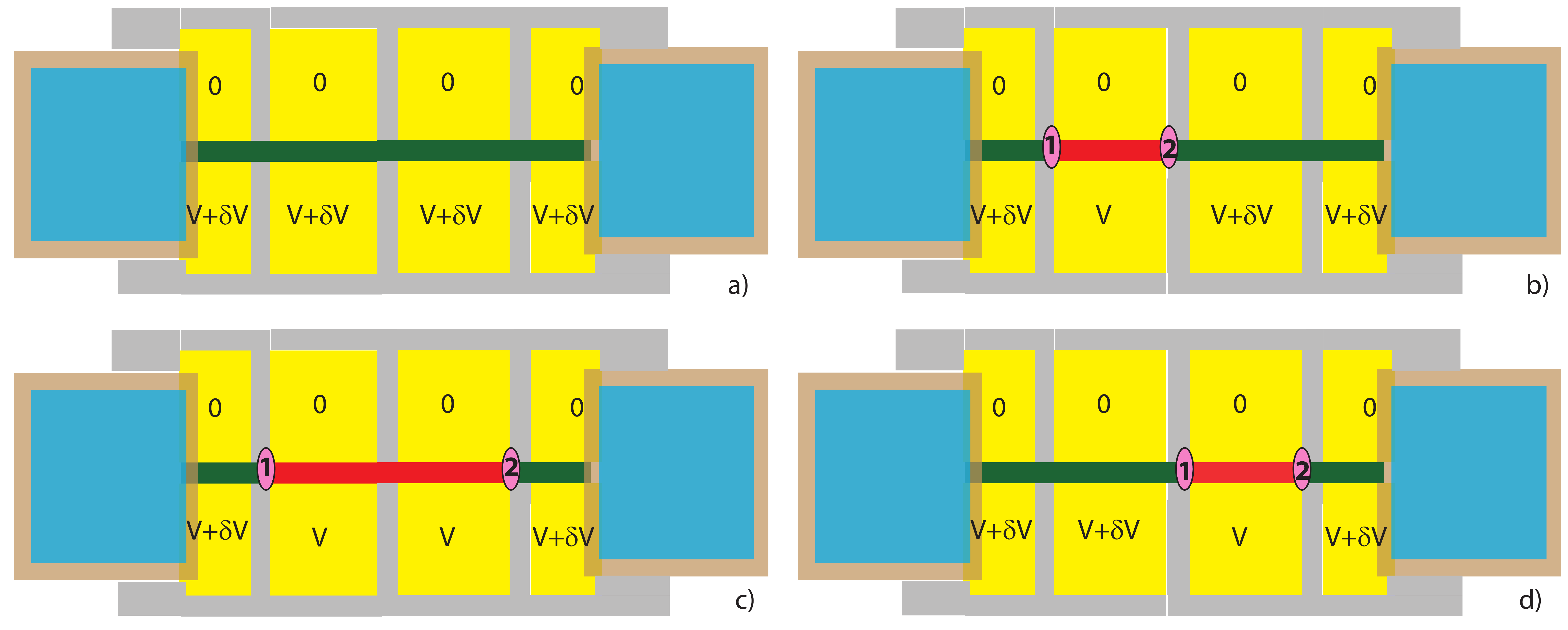}
\caption{Creating and moving a Majorana pair: a) Setting voltage differences between top and bottom gates to $V+\delta V$  yields trivial supercronductivity in all domain walls; b) setting voltage to $V$ on the second bottom gate  drives the system into the topological phase in
the domain wall above that gate and induces the Majorana modes at the ends of the domain wall; c) Setting the voltage to $V$ on the third bottom
gate extends the topological region to the domain wall above that gate
and moves one of the Majorana modes to a new boundary between topological and non-topological state; d) Setting the voltage to $V+\delta V$ on the second bottom gate moves the first of Majorana modes to the right. Blue areas are s-superconductors, yellow areas are top gates. Difference of voltages between two neighboring yellow gates defines the presence of domain wall and the type of the superconducting order parameter.
Red domain walls are in topological superconducting state, and green domain walls are non-topological superconductors. Grey areas correspond to voltage differences between neighboring gates insufficient to create a domain wall. }
\label{fig:Majorana_move}
\end{figure}

\begin{figure}
\includegraphics[width=0.8\textwidth]{MExch.pdf}
\caption{Exchanging a pair of Majorana modes using method of moving the Majorana pair.} \label{fig:exch}
\end{figure}

\begin{figure}
\includegraphics[width=0.9\textwidth]{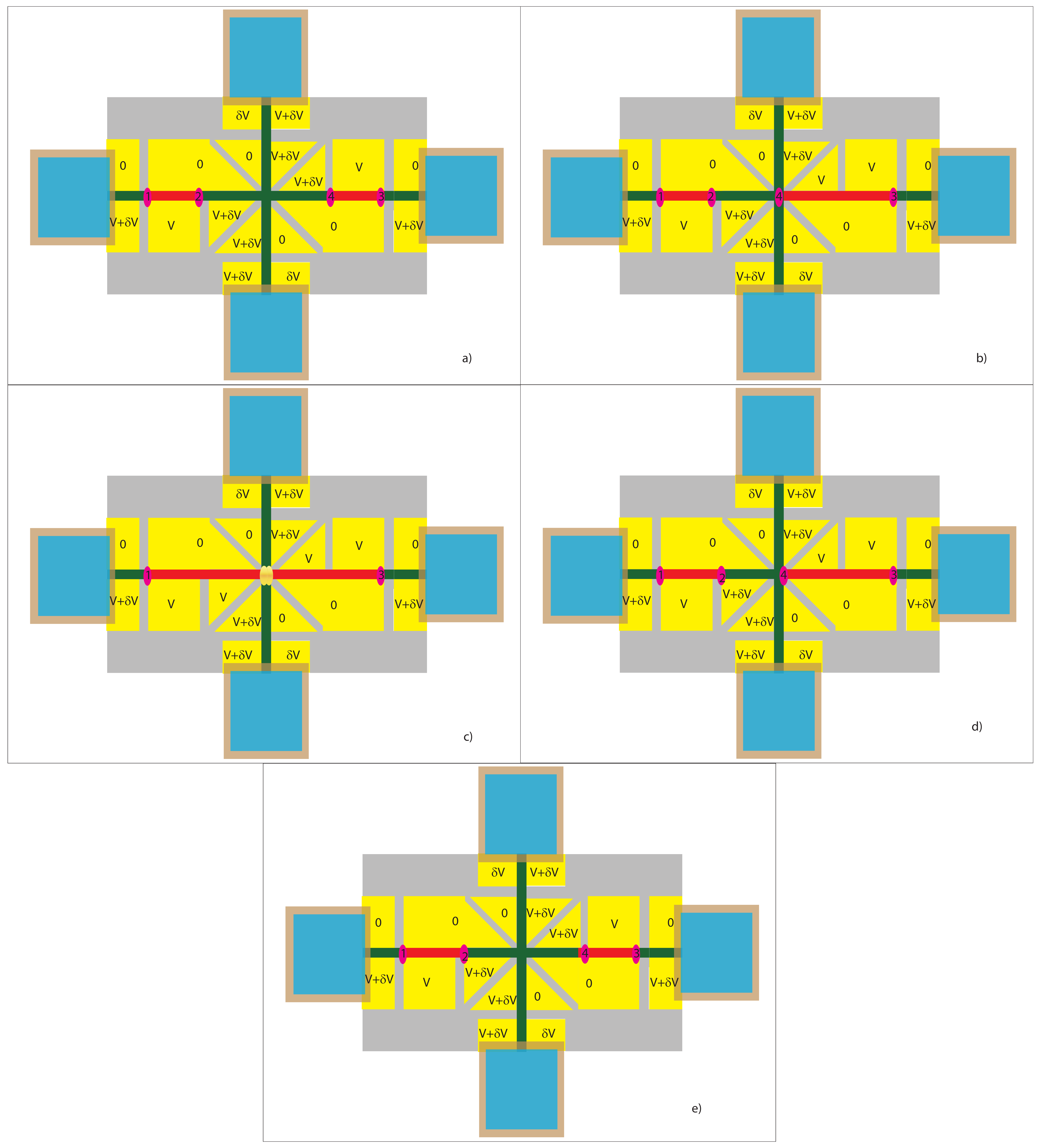}
\caption{Fusion and recreation of Majorana modes using method of moving the Majorana pair.} \label{fig:fuse}
\end{figure}

\begin{figure}
\includegraphics[width=0.7\textwidth]{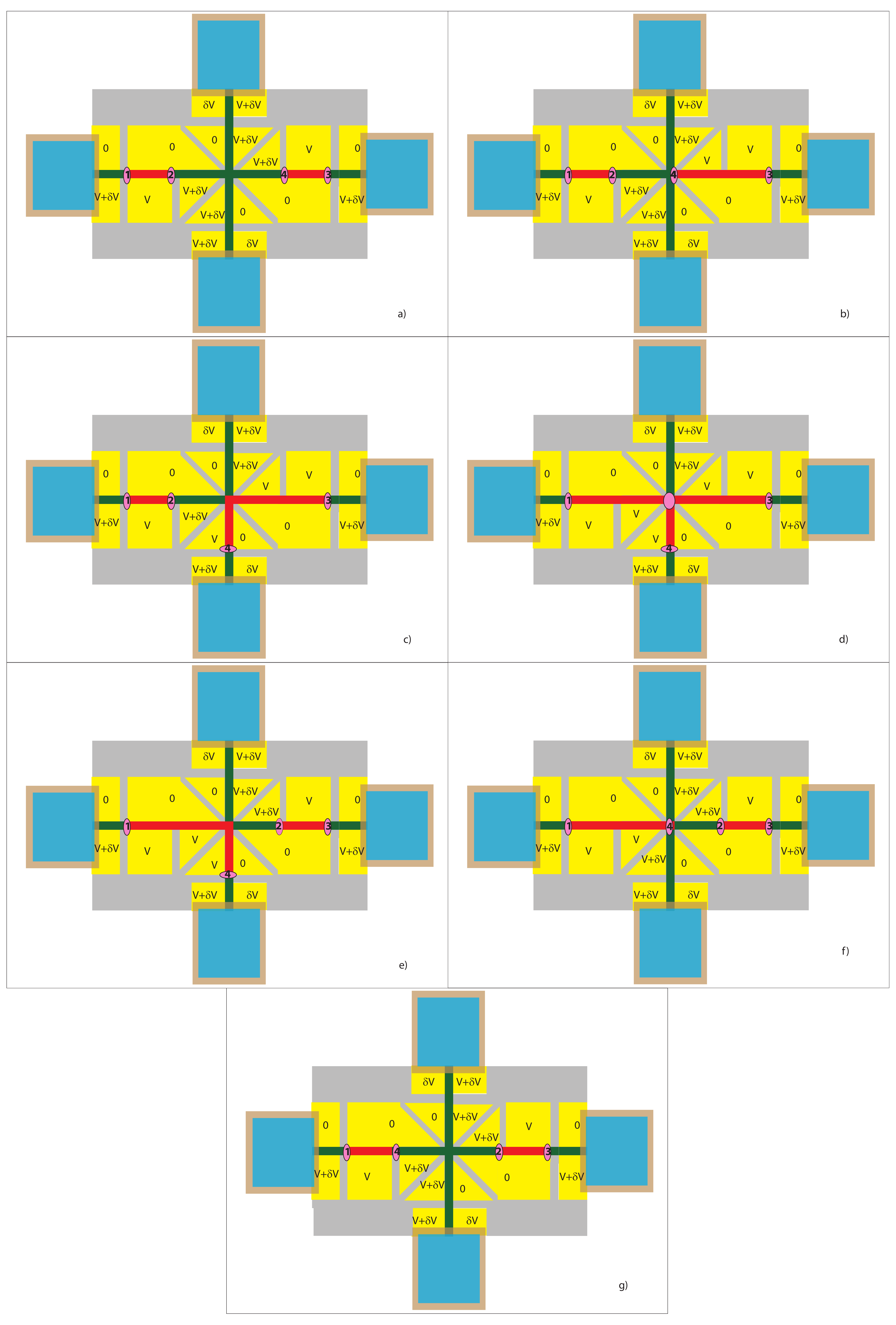}
\caption{Braiding Majorana modes achieved using a  method of moving the Majorana pair and a $T$-junction of domain walls in topological superconducting state. } \label{MBraid}
\end{figure}

\subsection{Control of Majorana modes}

Numerical simulations are performed for heterostructures studied in \cite{Kazakov2017} assuming $\Delta=0.1$ meV, $\gamma_R=0.44$ meV$\cdot$nm, and $\mu=32$ $\mu$eV.
We estimate that the voltage difference between the gates $V=V_1-V_2\sim 129$ mV corresponds to the topological condition $\hbar v/\ell < 0.3$ with Majorana fermions formed at the end of the chain, while additional voltage $\delta V\sim 1$ mV (total voltage difference $V+\delta V$) brings the system to the normal superconducting proximity state.

Thus, using electrostatic gates, we can move Majorana modes, and create and annihilate them. Furthermore, reduction of the difference  of voltages on electrostatic gates on the
sides of the domain wall area to a voltage below 10 meV (in theory, making it zero) erases the domain wall altogether, and can also serve as an instrument in manipulating reconfigurable network of topological superconductors. Figs. \ref{fig:Majorana_move}, \ref{fig:exch}, \ref{fig:fuse} and \ref{MBraid} demonstrate  inducing, moving, exchange, fusion and braiding of Majorana modes. In these figures, blue areas are s-superconductors and yellow areas are top gates. Difference of voltages between two neighboring yellow gates defines the presence of domain wall and the type of the superconducting order parameter.
Red domain walls are in topological superconducting state, and green domain walls are non-topological superconductors, while grey areas correspond to voltage differences between neighboring gates insufficient to create a domain wall.
Braiding of Majorana fermions are achieved using a structure containing $T$-junction of domain walls in topological superconducting state, Fig. \ref{MBraid}. By moving Majorana modes,
 two pairs of such modes are brought to a T-junction as in panel d). Then a  $T$-junction link is cut  by increasing the voltage by $\delta V$ on the gate controlling that link. Gate voltages are then brought back to the initial configuration. We underscore that all manipulations are expected to be produced by voltage pulses. Calculated parameters and requirements for the scheme are realistic and feasible for experiments in near future.

We note that in the schemes Fig.\ref{fig:Majorana_move}-\ref{MBraid}, a superconducting pairing potential $\Delta$ is assumed spatially uniform in the domain wall areas. In real settings with superconducting contacts on the sides of the domain walls, the induced superconducting gap is expected to be spatially dependent, decreasing from the contact area into the sample. Spatially dependent $\Delta(y)$ will re-define boundaries between topological and non-topological superconducting regions. These boundaries, and Majorana modes residing at boundaries, can be moved with adjusted gate voltages, when applied gate voltage exceeds the critical value in an area with lower $\Delta$  but is smaller than the critical value
in the area closer to the contact.

\section{Conclusion}
In this work we considered Majorana modes in hybrid s-superconductor - filling factor $\nu=2$ quantum Hall ferromagnet domain wall system. We discovered that when the Fermi level is pinned to a gap between anticrossing
spin-orbit coupled edge states, the impurity disorder in short domain walls generates proximity-induced topological superconductivity and the Majorana zero modes. Thus, in this case not only topological superconductivity is disorder robust,
but it emerges exclusively due to impurity disorder. Hybrid structures of s-superconductor with fractional quantum Hall  edge states were suggested as possible realization of parafermions, which could bring such settings closer to fault-tolerant quantum computing. Quantum Hall ferromagnet domain walls at fractional filling factors proximity-coupled to s-type superconductor can also potentially produce parafermions, making studies of helical domain walls an important area of the field of topological quantum computing.

\vspace{0.5cm}
\textbf{Acknowledgements}\\
\begin{acknowledgments}
G.S., A.K., L.P.R., T.W. and Y.B.L-G. acknowledge support by the  Department of Defence Office of Naval Research Award N000141410339. T.W. was partially supported by the National Science Centre (Poland) through Grant No. DEC- 2012/06/A/ST3/00247 and by the Foundation for Polish Science through the IRA Program co-financed by EU within SG OP.
\end{acknowledgments}

%

\end{document}